\documentclass[fleqn,twoside]{article}

\newcommand{\bls}[1]{\renewcommand{\baselinestretch}{#1}}
\newcommand{\Abstract}[1]{\vskip 2mm \begin{center}
        \parbox{16.4cm}{\small\noi #1} \end{center}\medskip}
\newcommand{\noi}{\noindent}
\newcommand{\Title}[1]{\noi {\Large #1} \\}

\newcommand{\email}[2]{\footnotetext[#1]{e-mail: #2}}
\def\nn{\nonumber\\ {}}
\def\beq#1{\begin{equation}\label{#1}}
\def\eeq{\end{equation}}
\def\dys{\displaystyle}
\def\mst{\mathstrut}

\catcode`\@=11 \@addtoreset{equation}{section}\catcode`\@=12

\newcommand{\bear}[1]{\begin{eqnarray}\label{#1}}
\newcommand{\ear}{\end{eqnarray}}

\newcommand{\R}{{\bf R}}

\renewcommand{\nn}{\nonumber}

\bls{0.957}

\begin{document}

\Title  {\uppercase{MULTIDIMENSIONAL BLACK HOLE SOLUTIONS \\\\
IN MODEL WITH PERFECT FLUID}}

{\bf V.D. Ivashchuk,$^{b,1}$ V.N. Melnikov$^{b,2}$ and A.B.
Selivanov$^{a,3}$}

\medskip{\protect
\begin{description}\itemsep -1pt
     \item[$^a$]{\it Institute of Gravitation and Cosmology,Peoples
Friendship \\
      University of Russia, 6 Miklukho-Maklaya St., Moscow 117198, Russia}

     \item[$^b$]{\it Centre for Gravitation and Fundamental\\
     Metrology, VNIIMS, 3-1 M.Ulyanovoy St., Moscow 117313,
     Russia}

\end{description}}

\Abstract { A family of black-hole solutions in the model with
1-component perfect fluid is obtained.  The metric of any solution
contains $(n -1)$ Ricci-flat ``internal space'' metrics and for
certain equations of state ($p_i = \pm \rho$) coincides with the
metric of black brane (or black hole) solution in the model with
antisymmetric form. Certain examples (e.g. imitating $M2$ and $M5$
black branes) are considered. The post-Newtonian parameters
$\beta$ and $\gamma$ corresponding to the 4-dimensional section of
the metric are calculated.}

\email 1 {ivas@rgs.phys.msu.su} \email 2
{melnikov@rgs.phys.msu.su} \email 3 {seliv@rgs.phys.msu.su}

\section{Introduction}

Black brane solutions (see, for example, \cite{IMtop} and
references therein) defined on product  manifolds $\R \times M_0
\times \ldots \times M_n$ are widely interested now. The
solutions appear in the models with antisymmetric forms and
scalar fields. These and more general $p$-brane cosmological and
spherically symmetric solutions are usually obtained by the
reduction of the field equations to the Lagrange equations
corresponding to Toda-like  systems \cite{IMJ}. An analogous
reduction for the models with multicomponent perfect fluid was
done earlier in \cite{IM5}. For cosmological-type models with
antisymmetric forms without scalar fields any $p$-brane is
equivalent to  an anisotropic perfect fluid with the equations of
state:
\beq{0} p_i = - \rho  \quad {\rm or} \quad p_i = \rho,
\eeq
when the manifold $M_i$ belongs or does not belong to the
brane worldvolume, respectively (here $p_i$ is the pressure in
$M_i$  and $\rho$ is the density).

The aim of this paper is to find explicitly the analogues of
black brane (or black hole) solutions in the model with
1-component perfect fluid  and extend it to more general
equations of state.

The paper is organized as follows. In Section 2 the model is
formulated. In Section 3 the general black hole solutions are
presented. Section 4  deals with certain analogues of black brane
solutions, e.g. the Reissner-Nordstr\"{o}m solution and $M2$- and
$M5$- black brane solutions. In Section 5 the post-Newtonian
parameters for the 4-dimensional section of the metric are
calculated.

\section{Model}

In this paper we consider a family of spherically symmetric
solutions to Einstein equations with a perfect-fluid matter source

\beq{1.1} R^M_N - \frac{1}{2}\delta^M_N R = k T^M_N \eeq defined
on the manifold

\beq{1.2}
\begin{array}{l}
M = {\R}_{.}\times (M_{0}=S^{d_0}) \times (M_1 = {\R})
 \times \ldots \times M_n,\\ \qquad ^{radial
    \phantom{p}}_{variable}\quad^{spherical}_{variables}\quad\qquad^{time}
\end{array}
\eeq with the block-diagonal metrics \beq{1.2a} ds^2= e^{2\gamma
(u)} du^{2}+\sum^{n}_{i=1} e^{2X^i(u)} h^{(i)}_{m_in_i }
dy^{m_i}dy^{n_i }. \eeq

Here $\R_{.} = (a,b)$ is interval. The manifold $M_i$ with the
metric $h^{(i)}$, $i=1,2,\ldots,n$, is a Ricci-flat space of
dimension $d_{i}$:

\beq{1.3} R_{m_{i}n_{i}}[h^{(i)}]=0, \eeq and $h^{(0)}$ is
standard metric on the unit sphere $S^{d_0}$

\beq{1.4} R_{m_{0}n_{0}}[h^{(0)}]=(d_0-1)h^{(0)}_{m_{0}n_{0}},
\eeq $u$ is radial variable, $\kappa$ is the gravitational
constant, $d_1 = 1$ and $h^{(1)} = -dt \otimes dt$.

The energy-momentum tensor is adopted in the following form
\beq{1.5} (T^{M}_{N})= {\rm diag}(-{\hat{\rho}},{\hat p}_{0}
\delta^{m_{0}}_{k_{0}}, {\hat p}_{1}
\delta^{m_{1}}_{k_{1}},\ldots , {\hat p}_n
\delta^{m_{n}}_{k_{n}}), \eeq where $\hat{\rho}$ and $\hat p_{i}$
are "effective"  density and pressures, respectively, depending
upon the radial variable $u$. We also impose the following
equation of state \beq{1.7} {\hat
p}_i=\left(1-\frac{2U_i}{d_i}\right){\hat{\rho}}, \eeq where
$U_i$ are constants, $i= 0,1,2,\ldots,n$.

The physical density and pressures are related to the effective
("hat") ones by formulas \beq{1.7a} \rho = - {\hat p}_1, \quad
p_u = - \hat{\rho}, \quad p_i = \hat{p}_i, \quad (i \neq 1). \eeq

In what follows we put $\kappa =1$ for simplicity.

\section{Black hole solutions}

We will make some natural assumptions: \beq{2.1}
\begin{array}{l}
1^{o}.\quad U_0 = 0 \Leftrightarrow \hat p_0 = \hat\rho,\\
2^o.\quad U_1 = 1 \Leftrightarrow \hat p_1 = -\hat\rho,\\
3^o.\quad (U,U)  = U_i G^{ij}U_j > 0,
\end{array}
\eeq where
\beq{2.2a}
G^{ij}=\frac{\delta^{ij}}{d_i} + \frac{1}{2-D},
\eeq
are components of the matrix inverse to the matrix of the
minisuperspace metric \cite{IMZ} \beq{2.2} (G_{ij}) = (d_i
\delta_{ij} - d_i d_j), \eeq and $D=1+\sum\limits_{i=0}^n {d_i}$
is the total dimension. (It may be proved that the restriction $3^o$
follows from from the first ones: $1^o$ and $2^o$.)

It follows from (\ref{1.7}) and restriction $1^o$ that

\beq{2.1a} \rho = \hat{\rho}, \quad p_u = - \rho, \quad p_0 =
\rho. \eeq

Under the relations  (\ref{1.7}) and (\ref{2.1}) imposed we
obtained the following black-hole solutions to the Einstein
equations (\ref{1.1}):

\bear{12} \nn ds^{2} = J_{0}\left( \dys\frac{\mst
dr^{2}}{1-\frac{2\mu}{r^{d}}} + r^{2} d \Omega^2_{d_0} \right) -
J_1\left(1-\frac{2\mu}{r^{d}}\right)dt^{2}
+ \sum_{i=2}^{n} J_{i} h^{(i)}_{m_{i}n_{i}} dy^{m_{i}}dy^{n_{i}},\quad \\
\label{13} \rho=\dys\frac{\mst d^2 \nu^2 P(P+2\mu)}{2H^2 J_0
r^{2d_0}},\quad
\ear
that may be verified by analogy with the
$p$-brane solution \cite{IMJ} (the detailed treatment will be in
a separate publication). Here $d=d_0-1$, $d \Omega_{d_0}^2=
h^{(0)}_{m_{0}n_{0}} dy^{m_{0}}dy^{n_{0}}$ is spherical element,
the metric factors

\beq{2.3} J_{i} = H^{-2\nu^{2}U^{i}}, \quad H =
1+\frac{P}{r^{d}}; \eeq $P >0$, $\mu >0$ are integration
constants and

\bear{2.4} U^{i} = G^{ij}U_{j}  = \frac{U_i}{d_i} + \frac{1}{2-D}
\sum_{j=0}^{n}U_j ,
\\ \label{2.4a} \nu = (U,U)^{-1/2}. \ear

Using (\ref{2.4}) and the first assumption from (\ref{2.1})
one can rewrite (\ref{12}) as follows

\bear {12a}
ds^{2} = J_{0} \Biggl[ \dys\frac{\mst
dr^{2}}{1-\frac{2\mu}{r^{d}}} + r^{2} d \Omega^2_{d_0} -
H^{-2\nu^{2}} \left(1-\frac{2\mu}{r^{d}}\right)dt^{2}
\sum\limits_{i=2}^{n} H^{-2\nu^{2}U_i/d_i} h^{(i)}_{m_{i}n_{i}}
dy^{m_{i}}dy^{n_{i}} \Biggr].\quad  \ear

\section{Imitation of black brane solutions}

Here we consider certain examples of solutions
with metrics of charged black hole and $M$-branes.

\subsection{Reissner-Nordstr\"{o}m solution}

Let us consider the  $4$-dimensional space-time manifold $\R \times
S^2 \times \R$. The metric and the density  from
(\ref{12a}) and (\ref{13}) read

\bear{3.1}  ds^2 = H^2 \left( \dys\frac{\mst
dr^2}{1-\frac{2\mu}{r}} + r^2 d\Omega^2_2 \right)-
H^{-2}(1-\frac{2\mu}{r}) dt^2,\ \\ \rho = \displaystyle\frac{\mst
P(P+2\mu)}{H^4 r^4}.\
\ear
By changing the variable $r = r' - P$
we obtain a standard Reissner-Nordstr\"{o}m metric with the
charge squared $Q^2 = P(P+2\mu)$ and the gravitational radius $GM
= P+\mu$

\subsection{Analogues of $M$-brane solutions.}

\quad Here we consider the case $D = 11$ and ${n=3}$.

{\bf M2 black brane}. For $U_2 = d_2 =2$, $U_3 = 0$ we get from
(\ref{12a}):

\bear{3.3}  ds^2 = H^{\frac{1}{3}} \Biggl[  \dys\frac{\mst
dr^{2}}{1-\frac{2\mu}{r^{d}}} + r^{2} d \Omega^2_{d_0} - H^{-1}
(1-\frac{2\mu}{r^d}) dt^2 + H^{-1} h^{(2)}_{m_{2}n_{2}}
dy^{m_{2}}dy^{n_{2}} + h^{(3)}_{m_{3}n_{3}}
dy^{m_{3}}dy^{n_{3}}\Biggr].\quad \ear
This relation corresponds to the metrics of the electric $M2$
black brane solution in 11-dimensional supergravity \cite{St,CT}.
The density (\ref{13}) has the following form:

\beq{3.4} \rho = \frac{ d^2 P(P+2\mu)}{4 H^{7/3} r^{2d_0}}.
\eeq

{\bf M5 black brane}. Let us consider another example in $D=11$
with $U_2 = d_2 = 5, U_3 =0$).
The metric reads
\bear{3.5} ds^2 = H^{\frac{2}{3}} \Biggl[  \dys\frac{\mst
dr^{2}}{1-\frac{2\mu}{r^{d}}} + r^{2} d \Omega^2_{d_0} - H^{-1}
(1-\frac{2\mu}{r^d}) dt^2 + H^{-1} h^{(2)}_{m_{2}n_{2}}
dy^{m_{2}}dy^{n_{2}} + h^{(3)}_{m_{3}n_{3}}
dy^{m_{3}}dy^{n_{3}}\Biggr].\quad \ear
and the density is as follows

\beq{3.6} \rho = \frac{ d^2 P(P+2\mu)}{4 H^{8/3} r^{2d_0}}. \eeq

The metric coincides with that of well-known $M5$  solution
\cite{St,CT}.

\section{Physical parameters}

\subsection{Gravitational mass and post-Newtonian parameters}

Here we put $d_0  =2 \ (d =1)$. Let us
consider the 4-dimensional space-time section of the metric
(\ref{12a}). Introducing a new radial variable by the relation:

\beq{3.7} r = R \left(1 + \frac{\mu}{2 R}\right)^2, \eeq we
rewrite the $4$-section in the following form:

\bear{3.8} ds_{(4)}^2 = g^{(4)}_{\mu \mu'} dx^{\mu} dx^{\mu'} =
H^{- 2\nu^2 U^0}\times  \left[ - H^{- 2\nu^2 }
\left(\frac{1-\frac{\mu}{2 R}}{1+\frac{\mu}{2 R}}\right)^2 dt^2 +
\left( 1+\frac{\mu}{2 R} \right)^4 \delta_{ij} dx^i dx^j \right]
, \ear $i,j = 1,2,3$. Here $R^2 = \delta_{ij} x^i x^j$.

The parametrized post-Newtonian (Eddington) parameters are
defined by the well-known relations

\bear{3.9}
g^{(4)}_{00} = - (1-2V+2\beta V^2) + O (V^3), \\
\label{3.10} g^{(4)}_{ij} = \delta_{ij} (1 + 2 \gamma V) + O
(V^2), \ear $i,j = 1,2,3$. Here $V=\frac{GM}{R}$ is
the Newtonian potential, $M$ is a gravitational mass and $G$ is
the gravitational constant. From (\ref{3.8})-(\ref{3.10}) we obtain:

\beq{3.12} GM = \mu + \nu^2 P (1 +U^0) \eeq and \bear{3.13}
\beta - 1= \frac{\nu^2 P (P + 2 \mu)}{2 (GM)^2} (1 + U^0), \\
\label{3.13a} \gamma - 1= - \frac{\nu^2 P}{GM} (1 + 2U^0), \ear

For fixed $U_i$ the parameter $\beta$ is proportional to the  ratio
of two physical parameters:  the perfect fluid density parameter
$|A|= \frac12 \nu^2 P (P + 2 \mu)$,  and the gravitational radius
squared $(GM)^2$.

\subsection{Hawking temperature}

The Hawking temperature of the black hole may be calculated using the
relation from \cite{York} and has the following form:
\beq{5.79} T_H =
\frac{d}{4 \pi (2 \mu)^{1/d}} \left(\frac{2 \mu}{2 \mu +
P}\right)^{\nu^2}. \eeq

\section{Conclusions}

We have obtained a family of black-hole solutions in the model
with 1-component perfect fluid with the equation of state
(\ref{1.7}) and the relations (\ref{2.1}) imposed.
The metric of the solutions  contains  $(n -1)$
Ricci-flat ``internal'' space metrics. For certain equation of
state (with $p_i = \pm \rho$) the metric of solution may coincide
with that of black brane (or black hole) solution (in the model
with antisymmetric forms without dilatons). Here we suggested
certain examples imitating  4-dimensional charged black hole and
$M 2$, $M 5$ black brane solutions in $D=11$ supergravity.

Here we have calculated the post-Newtonian parameters $\beta$ and
$\gamma$ corresponding to the 4-dimensional section of the
metric.  The parameter $\beta$ is written in terms of ratios of
the physical parameters: the perfect fluid parameter $|A|$ and
the gravitational radius squared $(GM)^2$.

{\bf Acknowlegments}

This work was supported in part by the Russian Ministry of
Science and Technology, Russian Foundation for Basic Research and
Project SEE.

\end{document}